\documentclass[12pt]{article}
\usepackage{epsfig}
\usepackage{float}
\usepackage{graphicx}
\begin{document}
\rightline{NKU-2016-SF2}
\bigskip

\newcommand{\be}{\begin{equation}}
\newcommand{\ee}{\end{equation}}
\newcommand{\noi}{\noindent}
\newcommand{\ra}{\rightarrow}
\newcommand{\bib}{\bibitem}
\newcommand{\refb}[1]{(\ref{#1})}

\newcommand{\bff}{\begin{figure}}
\newcommand{\eff}{\end{figure}}


\begin{center}
{\Large\bf Bardeen-de Sitter   black holes}

\end{center}
\hspace{0.4cm}
\begin{center}
Sharmanthie Fernando \footnote{fernando@nku.edu} \\
{\small\it Department of Physics, Geology \& Engineering Technology}\\
{\small\it Northern Kentucky University}\\
{\small\it Highland Heights}\\
{\small\it Kentucky 41099}\\
{\small\it U.S.A.}\\

\end{center}

\begin{center}
{\bf Abstract}
\end{center}

In this paper we present a regular black hole with a positive cosmological constant. The regular black hole considered is the well known Bardeen black hole and it is a solution to the Einstein equations coupled to non-linear electrodynamics with a magnetic monopole. The paper discuss the properties of the Bardeen-de Sitter black hole. We have computed the grey body factors and partial absorption cross sections for massless scalar field impinges on this black hole with the third order WKB approximation. A detailed discussion on how  the behavior of the grey body factors depend on the parameters  of the theory such as the 
mass, charge and the cosmological constant is given. Possible extensions of the work is discussed at the end of the paper.

\hspace{0.7cm}



\section{Introduction}

There are many observations that have verified that the present day universe is undergoing accelerated expansion \cite{perl}\cite{riess}\cite{sper}\cite{sel}. The common wisdom is that there is some unknown ``dark energy'' that is causing this accelerated expansion. There are various models that have been proposed  in order to explain dark energy, but the cosmological constant seems to be the most popular of all. When a positive cosmological constant is introduced to general theory of relativity, the resulting space time geometry gain a positive curvature asymptotically. Here in this paper we study the Bardeen black hole with a positive cosmological constant. Hence we name this black hole as Bardeen-de Sitter black hole. This is one of few regular black holes with a non-flat asymptotic geometry.

Usually, a black hole has a horizon as well as a singularity inside the horizon where the curvature scalar goes to infinity. However, there are some black holes which do not have a singularity at the origin. Such black holes are called ``regular'' black holes.   Bardeen in 1968 obtained a black holes solution without a singularity and is now well known as the Bardeen black hole \cite{bardeen}. Bardeen black hole spacetime satisfy the weak energy condition but does not satisfy the strong energy condition. The Einstein's tensor for the Bardeen black hole is non-zero. 30 years later since Bardeen presented his solution, Ay\'{o}n-Beato and Garc\'{i}a proved that the Bardeen black hole can be interpreted as a gravitationally collapsed magnetic monopole arising in a specific form of non-linear electrodynamics \cite{gar1}. The corresponding Lagrangian for the suggested non-linear electrodynamics was presented. Hence the stress energy tensor of the non-linear electrodynamics given worked as the source for the Einstein's field equations.
Bardeen black hole has attracted lot of attention in the recent past. Stability properties of the Bardeen black hole and other regular black holes were studied by Moreno and Sarbach \cite{sar}. Quasinormal modes of Bardeen black holes have been studied by several authors \cite{fernando1} \cite{lemos} \cite{ulhoa}. Profiled spectral lines generated by Keplerian discs orbiting around Bardeen black holes were addressed in \cite{jan}.  Thermodynamic quantities of the Bardeen black holes were studied by Man and Cheng \cite{man}. The geodesic structure of test particle in Bardeen space-time was studied by Zhou et.al. \cite{zhou}.

A black hole does not act like a perfect black body when waves impinged on it. Instead, a part of the wave will scatter back to infinity due to the gravitational potential of the black hole. Therefore, on can define reflection coefficient ($R$) and transmission coefficient ($T$) which depends on the frequency of the wave. The grey body factor, $\gamma_l(\omega)$ can be defined as the probability of the waves impinges on the black hole to be absorbed by the black hole. In terms of the transmission coefficient, the grey body factor, $\gamma_l(\omega) = |T|^2$.

Studying the absorption and scattering cross sections of waves by black holes has been done since 1970's. There are many works dedicated to calculate greybody factors, absorption cross sections in the literature. It is not possible to include all of the works related to grey body factors, but will mention some of the interesting works we found in the literature. Greybody factors for scalar fields emitted by higher-dimensional Schwarzschild-de Sitter black holes were studied by Kanti et.al. \cite{kanti1}. The temperature dependence of the absorption cross section for black holes in string theory is given in \cite{moura}. Greybody factors for topologically massless black holes for non-minimally coupled scalar field  is studied by Gonzalez et.al\cite{gon}. In \cite{myung}, absorption cross section in de Sitter space in three dimensions were studied. Massless scalar  field emission in 2+1 dimensional dilaton charged black holes were studied by Fernando \cite{fernando2}. Low energy greybody factors for fermions emitted by the Schwarzschild-de Sitter black hole was studied  in \cite{spo}. Black hole grey body factors and D-brane spectroscopy were studied by Maldesena and Strominger \cite{mal}.  Since we are studying the Bardeen black hole in this paper, we would like to mention works that have addressed absorption cross sections of such black holes: absorption cross section for the Bardeen black hole with $\Lambda =0$ was done in \cite{macedo}, \cite{caio1} \cite{caio2} and \cite{tosh1}.


\section{Introduction to the  Bardeen-de Sitter  black hole}

In this section, we will extend the work of  \cite{gar1} to include a cosmological constant to the theory. The corresponding action is given by,

\begin{equation} \label{action}
S = \int d^4x \sqrt{-g} \left[ \frac{(R - 2 \Lambda) }{16 \pi } - \frac{ 1}{ 4 \pi} {\cal L}(F) \right]
\end{equation}
Here, $R$ is the scalar curvature, and ${\cal L}(F)$ is a function of $F = \frac{1}{4}F_{\mu \nu} F^{\mu \nu}$ given by,
\begin{equation} \label{lag}
{\cal L} (F) = \frac{ 3}{ 2 s g^2} \left( \frac{ \sqrt{2 g^2 F}}{ 1 + \sqrt{ 2 g^2 F}} \right)^{\frac{5}{2}}
\end{equation}
The field strength of the non-linear electrodynamics, $F_{\mu \nu}$ is given by, 
\be
F_{\mu \nu} = 2 ( \bigtriangledown_{\mu} A_ {\nu} - \bigtriangledown_{\nu} A_ {\mu})
\ee
The parameter $s$ in the above equation is given by $s = \frac{ |g|}{ 2 M}$ where $g$ and $ M$ corresponds to the magnetic charge and the mass of the black hole.

The field equations of motion derived from the action in eq$\refb{action}$  is given by,

\begin{equation} \label{ein}
G_{\mu \nu} + \Lambda g_{\mu \nu} = 2 \left( \frac{ \partial {\cal L(F)}}{ \partial F} F_{\mu \lambda} F_{ \nu}^ {\lambda} - g_{\mu \nu} {\cal L(F)} \right)
\end{equation}
\begin{equation} \label{max1}
\bigtriangledown_{\mu} \left(\frac{ \partial {\cal L(F)}}{ \partial F}  F^{\nu \mu}\right)=0
\end{equation}
\be \label{max2}
\bigtriangledown_{\mu} \left( *  F^{\nu \mu}\right)=0
\ee
We will look for space-time solutions which are static and spherically symmetric  for the above set of equations with a metric as follows:
\begin{equation}
ds^2 = -f(r) dt^2 + f(r)^{-1} dr^2 + r^2 ( d \theta^2 + sin^2(\theta) d \varphi^2)
\end{equation}
where,
\begin{equation}
f(r) = 1 - \frac{2 m(r)} { r}
\end{equation}
For a spherically symmetric space-time, the only non-vanishing components for $F_{\mu \nu}$ are $F_{01}$ and $F_{23}$.  We will follow the same ansatz made in \cite{gar1} for the Maxwell field as,
\be 
F_{\mu \nu} = 2 \delta^{\theta}_{[ \mu} \delta^{\varphi}_{\nu ]} Z(r , \theta)
\ee
When this ansatz is substituted to eq.$\refb{max1}$, it can be integrated easily to obtain,
\be
F_{\mu \nu} = 2 \delta^{\theta}_{[ \mu} \delta^{\varphi}_{\nu ]} g(r) sin \theta
\ee
The eq.$\refb{max2}$, implies  d F = 0; Hence $g'(r) sin \theta dr  \wedge d \theta \wedge d \varphi =0$, which lead to the conclusion that $ g(r) = const = g$. Here $g$ is the magnetic monopole charge. Hence, the magnetic field strength is given by,
\begin{equation}
F_{23} =  2 g sin \theta
\end{equation}
and
\be \label{fvalue}
F = \frac{ g^2}{ 2 r^4}
\ee
By substituting the value of $F$ in eq.$\refb{fvalue}$ to eq.$\refb{lag}$ one can obtain,
\be \label{lag2}
{ \cal L(F)} = \frac{ 3 M g^2}{ ( r^2 + g^2)^{\frac{5}{2}}}
\ee
From the Einstein field equations, eq$\refb{ein}$, the $G_{tt}$ component can be written as,
\be \label{ein2}
G_{tt} + \Lambda g_{tt} = - 2 {\cal L(F)} g_{tt}
\ee
where,
\be \label{gtt}
G_{tt} = - 2 \left( 1 - \frac{ 2 m(r)}{r} \right) \frac{ m'(r)} {r^2}
\ee
By substituting eq$\refb{gtt}$ and eq$\refb{lag2}$ into eq$\refb{ein2}$, one get,
\be \label{final}
m'(r) -  \frac{\Lambda r^2}{2} = \frac{ 3 M g^2 r^2}{ ( r^2 + g^2)^{\frac{5}{2}}}
\ee
Integrating  eq$\refb{final}$ from $r$ to $\infty$, one get,
\be \label{final2}
 - m(r) + \frac{ \Lambda r^3}{6} +  constant = \int^{\infty}_{r}  \frac{ 3 M g^2 r^2}{ ( r^2 + g^2)^{\frac{5}{2}}}
 \ee
Here,
\be
constant = lim_{ r \ra \infty} \left( m(r) - \frac{ \Lambda r^3}{6} \right) = M
\ee
The last integral in eq$\refb{final2}$ leads to an exact result given by,
\be
\int^{\infty}_{r}  \frac{ 3 M g^2 r^2}{ ( r^2 + g^2)^{\frac{5}{2}}} = M -  \frac{ M r^3}{ ( g^2 + r^2)^{\frac{3}{2}}}
\ee
Finally, the function $m(r)$ is given by,
\be
m(r) = \frac{ \Lambda r^3}{ 6} +  \frac{ M r^3}{ ( g^2 + r^2) ^{\frac{3}{2}}}
\ee
Substituting the expression for $m(r)$ in to the $f(r)$, one get,
\be
f(r)  = 1 - \frac{ 2 M r^2}{ (r^2 + g^2)^{\frac{3}{2}}} -  \frac{\Lambda r^2}{3}
\ee

\subsection{Horizons of the black holes}

\begin{figure} [H]
\begin{center}
\includegraphics{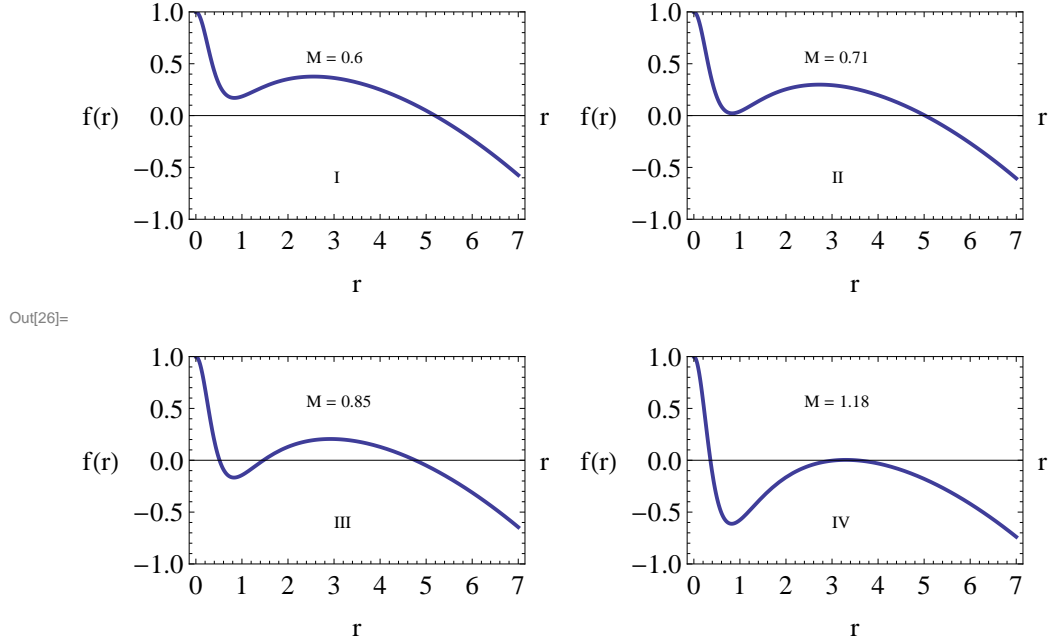}
\caption{Figure shows  $f(r)$ vs $r$ for $ \Lambda =0.086$ and $g =0.57$. The mass $M$ is varied to obtain different horizon structures.}
\label{fr}
 \end{center}
 \end{figure}

\begin{figure} [H]
\begin{center}
\includegraphics{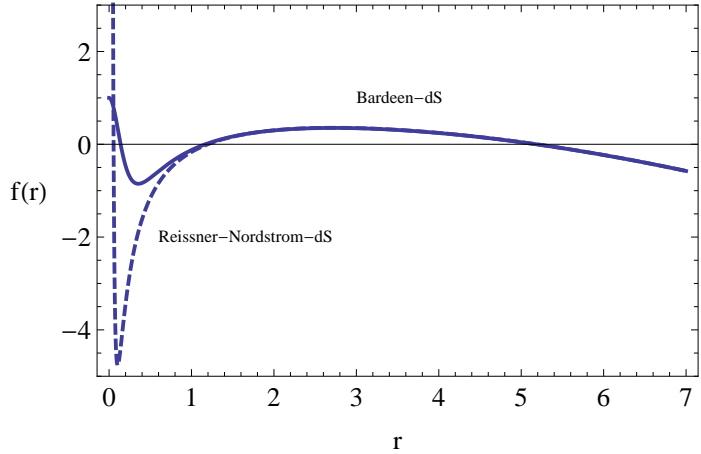}
\caption{Figure shows  $f(r)$ vs $r$ for Reissner-Nordstrom-de Sitter and Bardeen-de Sitter black holes. Here $ \Lambda =0.086, M =0.6$ and $g =0.25$}
\label{frboth}
 \end{center}
 \end{figure}

To determine the horizon structures, one can study the roots of $f(r) =0$.  $f(r)=0$ leads to the polynomial,
$$
 r^{10} \Lambda^2 + r^8 ( 3 g^2 \Lambda^2 - 6 \Lambda) + r^6( 3 \Lambda^2 g^4 - 18 g^2 \Lambda + 9) + r^4 ( g^6 \Lambda^2 - 18 g^4 \Lambda - 36 M^2 + 27 g^2)  
$$
\be
+ r^2 ( 27 g^4 - 6 g^6 \Lambda) + 9 g^6 =0
\ee
There can be three real roots, one real root or degenerate roots for the above equation.  Fig$\refb{fr}$ demonstrates different root structure for $f(r)$. The Bardeen-de Sitter black hole could have three horizons: inner horizon($r_i$), event horizon($r_h$) and the cosmological horizon($r_c$). It can have degenerate horizons when two of these merge. It can also have only one horizon (cosmological horizon)  for certain values of the  parameters of the theory. We have also plotted the function $f(r)$ for the Reissner-Nordstrom-de Sitter black hole and the Bardeen black hole in  the same figure which is given in Fig$\refb{frboth}$. The inner horizon is smaller for the Reissner-Nordstrom-de Sitter black hole 
compared to the Bardeen-de Sitter black hole. The event horizon is larger for the Reissner-Nordstrom-de Sitter black hole compared to the Bardeen-de Sitter black hole.

The Bardeen-de Sitter black hole is non-singular everywhere since all the scalar curvatures, $ R$, $ R_{\mu \nu} R^{ \mu \nu}$ and $ R_{\mu \nu \alpha \beta} R^{ \mu \nu \alpha \beta }$ are finite everywhere.  Only the electromagnetic invariant $ F = \frac{g^2}{ 2 r^4}$ has singular behavior at $r =0$. The Bardeen-de Sitter  black hole satisfy the weak energy condition.

Asymptotically, the metric function $f(r)$ behaves as,
\begin{equation}
f(r) \approx 1 - \frac{ 2 M}{r}  + \frac{3 M g^2}{ r^3} - \frac{ \Lambda r^2}{3}  + O \left( \frac{ 1}{ r^5} \right)
\end{equation}


\subsection{Extreme black holes}

Bardeen-de Sitter black hole could have degenerate horizons depending on the parameters $M, g$ and $\Lambda$.  Degenerate horizons occur when $ f(r)= f'(r)=0$. We will discuss three different types of degenerate black holes that may occur.

\subsubsection{ Nariai black holes}

When the cosmological and the event horizon merge, one get  Nariai black holes. This special case is represented by the Figure.(1) (IV). In these black holes, the degenerate horizon is given by,
\be \label{nariradius}
r^2_{Na}  =  \frac{ 1 + \sqrt{ 1 - 8 g^2 \Lambda} } { 2 \Lambda}
\ee
The corresponding mass is given by,
\be \label{nari}
M_{Nariai} =  \frac{ ( g^2 + r_{Na}^2 ) ^{5/2} \Lambda}{ 3 ( r_{Na}^2 - 2 g^2)}
\ee
In obtaining eq.$\refb{nari}$, we have substituted $r_{Na}$ to  $f'(r_{Na})=0$.

From eq$\refb{nariradius}$, one can see that the degenerate black holes exists only if $ 8 \Lambda g^2 < 1 $.

For a nearly extreme black hole, such as Nariai black hole,  one can approximate the function $f(r)$ as,

\begin{equation}
f(r) = \frac{ f''(r_{ex})}{2} ( r - r_c) ( r - r_h)
\end{equation}
Now, we shall introduce new coordinates,  $\chi$  and $ \psi$ as,
\begin{equation}
r = r_{ex} + \epsilon cos \chi
\end{equation}
\begin{equation}
t = \frac{ 2 \psi}{ \epsilon f''(r_{ex}) }
\end{equation}
Here $\epsilon$ is small. Hence, $\chi=0$ corresponds to $r = r_{ex} + \epsilon = r_c$ and $\chi = \pi$ corresponds to $r = r_{ex} - \epsilon = r_h$. The new coordinate  $\psi$ is   time like.
One can note  that $ f''(r_{ex}) < 0$ for the Nariai black hole   due to the nature of the function $f(r)$ at $ r = r_{ex}$. Now, after  substituting the new coordinates,  and taking the limit $ \epsilon \ra 0$, the metric simplifies to 
\be
ds^2 = \frac{ 2}{ f''(r_{ex})} \left(  sin^2\chi d \psi^2 - d \chi^2 \right) + r_{ex}^2 d \Omega^2
\end{equation}
The above geometry corresponds to $ dS_2 \times S^2$. The $dS_2$ has a positive scalar curvature,  
\be
 R_{dS_2}=    -f''(r_{ex})
 \ee
For the Bardeen-de Sitter  black hole,
\be \label{fdouble}
f''(r_{ex}) =  \frac{ 2 r_{ex}^2 \Lambda ( r_{ex}^2 - 4 g^2)}{ ( 2 g^4 + g^2 r_{ex}^2 - r_{ex}^4)}
\ee

\subsubsection{Cold black holes}

When the inner and the outer horizons coincides, one get cold black hole with zero temperature.  This special case is represented by Figure 1 (II). Similar to the Nariai black holes, at the degenerate horizons $f(r_{ex}) = f'(r_{ex}) =0$. Due to the nature of the function $f(r)$ at the horizons, $f''(r_{ex}) >0$.  The horizon of cold black holes are given by,
\be
r^2_{cold}  =  \frac{ 1 - \sqrt{ 1 - 8 g^2 \Lambda} } { 2 \Lambda}
\ee
The corresponding mass is given by,
\be
M_{cold} =  \frac{ ( g^2 + r_{cold}^2 ) ^{5/2} \Lambda}{ 3 ( r_{cold}^2 - 2 g^2)}
\ee
Two new coordinates $\psi$ and $\chi$ can be introduced as,
\begin{equation}
r = r_{ex} + \epsilon cosh \chi
\end{equation}
\begin{equation}
t = \frac{ 2 \psi}{ \epsilon f''(r_{ex}) }
\end{equation}
Once again $\epsilon$ is small and $\psi$ is a time-like coordinate. Once the new coordinates are substituted to the metric and taken the limit $\epsilon \ra 0$, the metric becomes,
\be \label{cold}
ds^2 = - \frac{ 2}{ f''(r_{ex})} \left(  sinh^2\chi d \psi^2 - d \chi^2 \right) + r_{ex}^2 d \Omega^2
\end{equation}
The above geometry represents $AdS_2 \times S^2$. The $AdS_2$ has the scalar  curvature $R_{cold} = -f''(r_{cold})$.


\subsubsection{Ultracold black holes}

In ultracold black holes, all three horizons coincides leading to the radius of the horizon,
\be
r_{ucold} = \frac{1}{\sqrt{2 \Lambda}} = 2 g
\ee
Here, $f = f'= f''=0$.
The mass of the ultracold black hole is given by,
\be
M_{ultracold} = \frac{ 25}{96} \sqrt{ \frac{ 5}{ 2 \Lambda}}
\ee
To understand the geometry of the ultracold black hole near the degenerate horizon, let us take the metric of the cold black hole  in eq$\refb{cold}$  and make an additional coordinate transformation as,
\be
\chi = \eta \sqrt{ \frac{ f''(r_{ex})}{2}} = \eta a
\ee
Here we have taken $a = \sqrt{ \frac{ f''(r_{ex})}{2}}$. When we substituted this to the metric of the cold-black hole in eq$\refb{cold}$, one get,
\be \label{ultracold}
ds^2 =  \left( \frac{sinh( \eta a)}{\eta a}\right)^2  d \psi^2 - d \chi^2 + r_{ex}^2 d \Omega^2
\end{equation}
Since $f''(r_{ex}) \ra 0$ for ultracold black holes, one can take the limit $f''(r_{ex}) \ra 0$ of eq$\refb{ultracold}$.  Since $\frac{ sinh( \eta a) }{ \eta a}  \ra 1$, the above metric simplifies to,
\begin{equation}
ds^2 =  - \eta^2 d \psi^2 +  d \eta^2 + r_{ucold}^2 d \Omega^2
\end{equation}
The  geometry given by the above metric  has the topology, $ R^2 \times S^2$. This is similar to the toplogy of the ultra-cold Reissner-Nordstrom- de Sitter black hole.

Nariai, cold and ultracold black holes  of Born-Infeld-de Sitter and quintessence black holes were studied by Fernando in \cite{fernando3}\cite{fernando7}\cite{fernando8}. Degenerate horizons of regular black holes were studied in \cite{mat2}.


\subsection{Thermodynamics}

The Hawking temperature of the Bardeen-de Sitter  black hole is given by,
\begin{equation}
T = - \frac{1}{4\pi} \left. \frac{ d g_{tt}}{d r} \right|_{r = r_h} = \frac{ 1 }{ 4 \pi} \left[ \frac{ 2 Mr_{h} ( r_{h}^2 - 2 g^2)}{ ( g^2 + r_{h}^2)^{5/2}  }- \frac{2 \Lambda r_h}{3} \right]
\end{equation}
Here, $r_h$ is the event horizon of the black hole which is a solution of $f(r)=0$. 


\section{Massless scalar perturbation of Bardeen-de Sitter  black holes}

In this section, we will introduce scalar perturbation by a massless field around the Bardeen black hole. A massless scalar field minimally coupled to gravity is described by the Klein-Gordon equation given by,
\begin{equation} \label{klein}
\frac{1}{\sqrt{-g}} \partial_{\mu} ( \sqrt{-g}  \partial^{\mu} \Phi ) =0
\end{equation}
For a monochromatic wave with frequency $\omega$ in a spherically symmetric space-time, one can introduce the ansatz for $\Phi$ as,
\begin{equation} \label{expansion}
\Phi =  \sum  e^{- i \omega t} Y_{l,m}(\theta,\phi) \frac{\xi(r)}{r} 
\end{equation}
With the ansatz given in eq$\refb{expansion}$, eq$\refb{klein}$ simplifies to a  Schr\"{o}dinger-type equation given by,
\begin{equation}
\frac{d^2 \xi(r)}{dr_{*}^2} + \left( \omega^2  -  V_s(r_*) \right) \xi(r) =0
\end{equation}
Here $V_s(r_*)$ is the scalar field potential given by,
\begin{equation} \label{pot}
V_s(r) =  \frac{l(l+1) f(r)}{r^2}  + \frac{f(r) f'(r) }{r} 
\end{equation}
and $r_*$ is the  well known ``tortoise'' coordinate given by,
\begin{equation}
dr_{*} = \frac{dr}{f(r)}
\end{equation}
When $r \rightarrow r_c$, $r_* \rightarrow \infty$ and when $r \rightarrow r_h$, $r_* \rightarrow - \infty$. The potential $V_s(r)$  goes to zero at  $ r = r_c, r_h$. 

From Fig$\refb{pot}$, when $l$ is increased, the height of the potential increases. This is the well-known suppression of the emission of modes with large $l$. Notice that when $l=0$, there is local minimum between the two horizons for the potential. 

When the magnetic charge $g$ is increased, the potential height becomes larger as given in Fig$\refb{potg}$ and there will be suppression of scalar field emission. On the other hand, when  $\Lambda$ is increased, the height of the potential decreases as  shown in Fig$\refb{potlambda}$ leading to an enhancement of emission of scalar fields.

We also plotted the potentials for the Bardeen-dS and the Schwarzschild-dS black holes in the same graph to compare in Fig$\refb{potboth}$. It is clear that the Bardeen-dS black hole has a higher potential. Hence the scalar fields are suppressed for the Bardeen-de Sitter black holes as compared with the Schwarzschild-de Sitter black holes.

\begin{figure} [H]
\begin{center}
\includegraphics{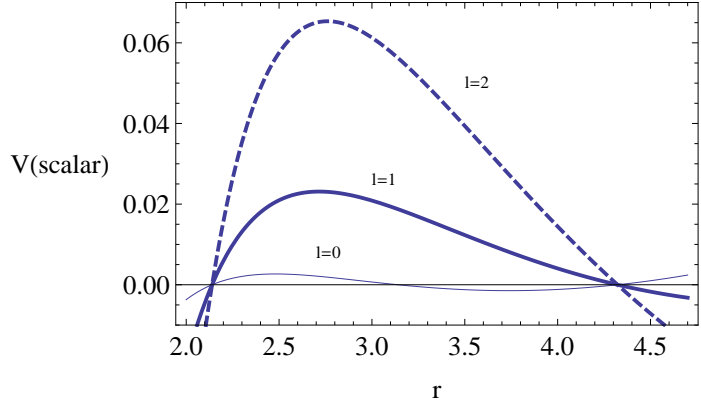}
\caption{The figure shows  $V_s(r)$ vs $r$ for $ \Lambda =0.086, M =1.03$ and $g =0.57$}
\label{pot}
 \end{center}
 \end{figure}

\begin{figure} [H]
\begin{center}
\includegraphics{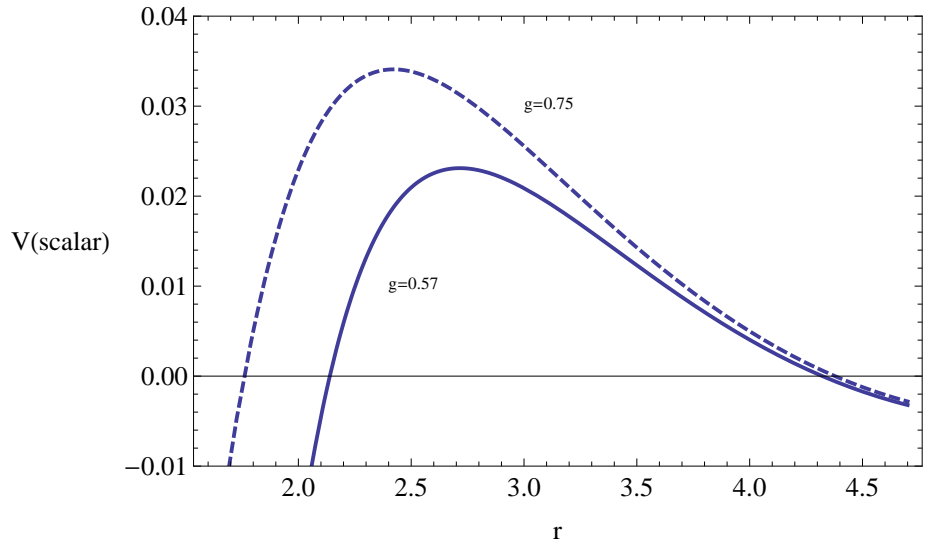}
\caption{The figure shows  $V_s(r)$ vs $r$ for $ \Lambda =0.086, M =1.03$ and $l=1$}
\label{potg}
 \end{center}
 \end{figure}
 
 \begin{figure} [H]
\begin{center}
\includegraphics{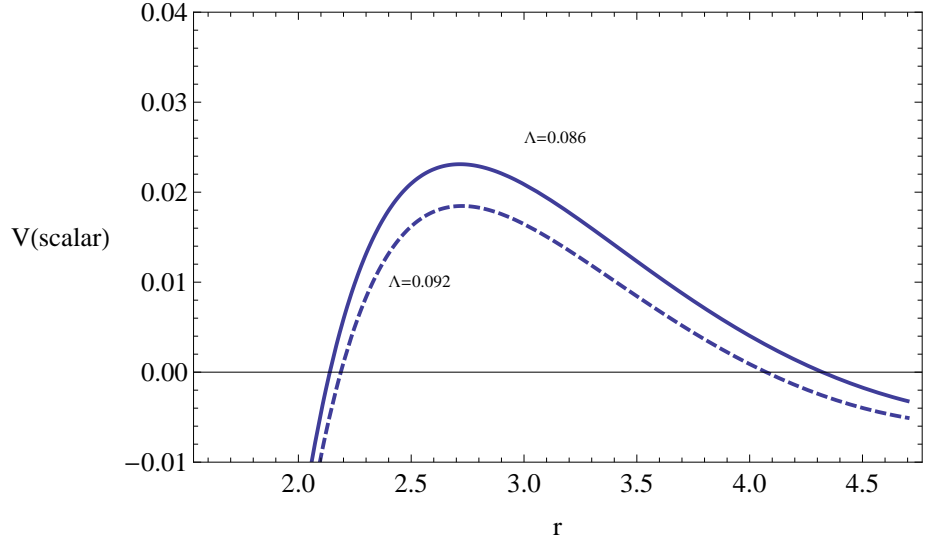}
\caption{The figure shows  $V_s(r)$ vs $r$ for $l=1, M =1.03$ and $g =0.57$}
\label{potlambda}
 \end{center}
 \end{figure}

\begin{figure} [H]
\begin{center}
\includegraphics{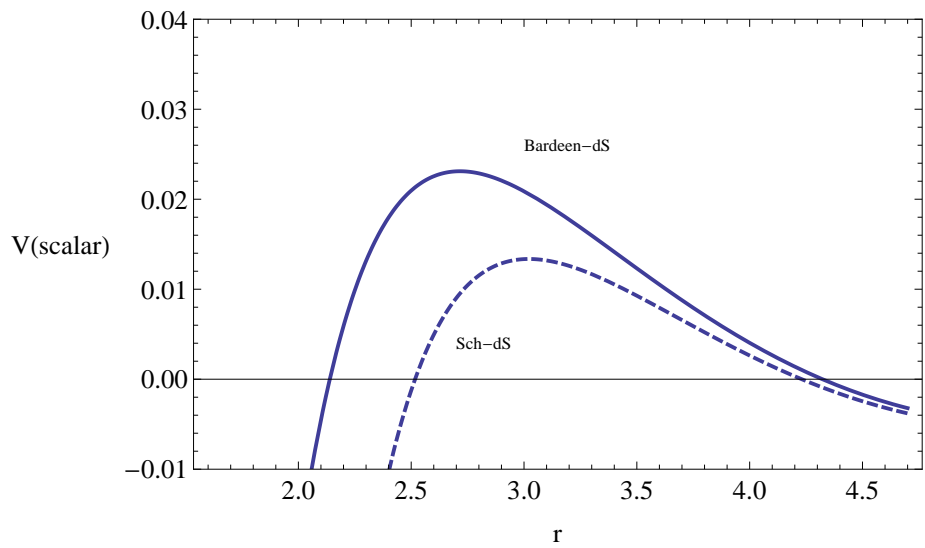}
\caption{The figure shows  $V_s(r)$ vs $r$ for the Bardeen-de Sitter and Schwarzschild-de Sitter black holes. Here, $ \Lambda =0.086, M =1.03, l=1$ and $g =0.57$}
\label{potboth}
 \end{center}
 \end{figure}


\section{ WKB approximation method to compute the greybody factors and partial absorption cross sections}

In this section we will describe the computation of reflection coefficients, transmission coefficients (greybody factors) and partial absorption cross sections  via the WKB approximation. 
Let us  assume a wave is coming from the past cosmological horizon which corresponds to $r* \ra \infty$ ( $r \ra r_c$). When the  wave reaches the black hole some of it will be reflected due to the gravitational potential and some of them will be transmitted. The reflected and the transmitted waves can be represented as the following:

\be
\xi(r_*) =  T(\omega) e^{ - i \omega r_*} \hspace{1 cm} r_* \ra - \infty( r \ra r_h)
\ee
\be
\xi(r_*) =   e^{ - i \omega r_*}    +   R(\omega) e^{  i \omega r_*}  \hspace{1 cm} r_* \ra   +\infty( r \ra r_c)
\ee
Here, $R(\omega)$ and $T(\omega)$ are the reflection  and transmission coefficient respectively. They are related by,
\be
|R(\omega)|^2 + |T(\omega)|^2 = 1
\ee
The greybody factor $\gamma_l(\omega)$   and the partial absorption cross section of the wave for a given frequency $\omega$ and $l$ is given by,
\be
\gamma_l(\omega) = |T(\omega)|^2
\ee
\be
\sigma_l = \frac{ \pi ( 2 l + 1)}{ \omega^2} |T(\omega)|^2
\ee

\subsection{WKB method}

In this section, we will describe the calculation of $R(\omega)$ and $T(\omega)$ using the WKB approximation. If  $r_0$ is the value of $r$ where the potential $V_s(r)$ is the maximum, then depending on the relation between $\omega$ and $V_s(r_0)$, there are three cases to consider:\\

\noindent
(a)  $\omega^2 << V_s(r_0)$: Here the transmission coefficient is close to zero and the reflection coefficient is almost equal to one.\\

\noindent
(b) $\omega^2 >> V_s(r_0)$: Here the transmission coefficient is close to one and the reflection coefficient is almost equal to zero.\\

\noindent
(c) $\omega^2$ is of the same order as $V_s(r_0)$: This is the case where we will focus on. It is known that the WKB approximation has high accuracy when $\omega^2 \approx V_s(r_0)$.\\

Now, let us describe the WKB method developed by Shutz and Will \cite{will} and Iyer and Will\cite{will2}: in this approximation, the reflection coefficient $R(\omega)$  is given by,
\be
R(\omega) = \left( 1 + e^{ - 2 \pi i \alpha }\right)^{-\frac{1}{2}} 
\ee
and
\be
|T(\omega)|^2 = 1 - |R(\omega)|^2
\ee
Here,
\be
\alpha =  \frac{ i ( \omega^2 - V_s(r_0))} { \sqrt{ - 2 V''_s(r_0)}}  - \Gamma_2 - \Gamma_3
\ee
In the above equation, $V_s''(r_0)$ is the value of the second derivative of $V_s(r)$ at $r_0$. $\Gamma_2$ and $\Gamma_3$  are the second and third order corrections to the WKB formula beyond the first order approximation.  $\Gamma_2$ and $\Gamma_3$ are given by,

\begin{equation}
\Gamma_2 = \frac{1}{(2 Q_0^{(2)})^{1/2}} \left[\frac{1}{8} \left[ \frac{Q_0^{(4)}}{Q_0^{(2)}} \right] ( \frac{1}{4} + \alpha^2 ) -\frac{1}{288} \left[\frac{Q_0^{(3)}}{Q_0^{(2)} } \right]^2 ( 7 + 60 \alpha^2 ) \right]
\end{equation}
$$
\Gamma_3= \frac{ n + \frac{1}{2}}{2 Q_0^{(2)}} \left[ \frac{5}{6912} \left[ \frac{ Q_0^{(3)}}{Q_0^{(2)}} \right]^4 ( 77 + 188 \alpha^2) - \frac{1}{384} \left[ \frac{(Q_0^{(3)})^2 Q_0^{(4)} }{ (Q_0^{(2)})^3} \right] ( 51 + 100 \alpha^2) \right.
$$
\begin{equation} 
\left. + \frac{1}{2304} \left[ \frac{Q_0^{(4)}}{Q_0^{(2)}} \right]^2 ( 67 + 68 \alpha^2) + \frac{1}{288} \left[ \frac{(Q_0^{(2)})^3 Q_0^{(5)}}{(Q_0^{(2)})^2} \right] ( 19+ 28 \alpha^2) 
-\frac{1}{288} \left[ \frac{Q_0^{(6)}}{ Q_0^{(2)} } \right] 
( 5 + 4 \alpha^2)\right] 
\end{equation}
Where,
\be
Q_0 = \omega^2 - V_s(r)
\ee
\be
Q_0^{(n)} = \frac{ d^nQ_0}{dr_*} |_{ r_* = r*(r_0)}
\ee

The WKB method to compute reflection coefficients, transmission coefficients and partial greybody factors have been employed in various papers:  greaybody factors of black holes in braneworld have been computed by WKB method in \cite{tosh2}. The WKB approach was also used to compute reflection coefficients of zero temperature superconductors by Konoplya and Zhidenko in \cite{kono5}. Same two authors used 6th order WKB approximation to calculate transmission coefficients of scattering of wormholes \cite{kono7}

\begin{figure} [H]
\begin{center}
\includegraphics{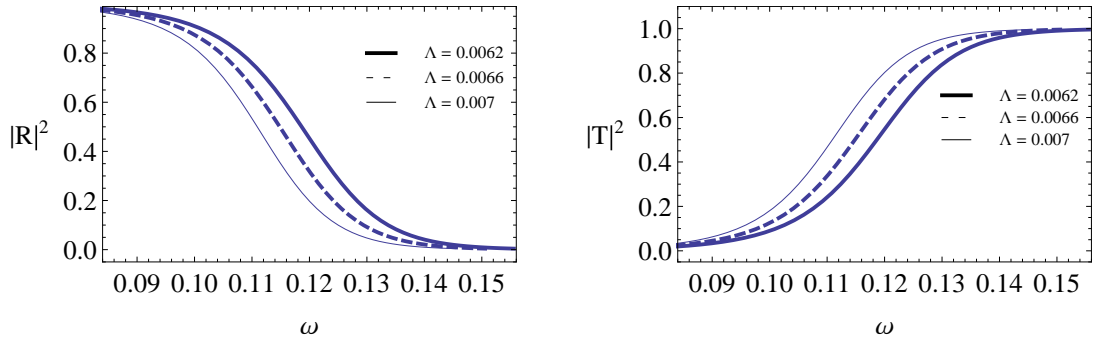}
\caption{The figure shows  $T(\omega)^2$  and $R(\omega)^2$ vs $\omega$ for $g = 1, M = 3$ and $l=2$. The value of $\Lambda$ takes the values $0.0062,0.0066, 0.007$.}
\label{translambda}
 \end{center}
 \end{figure}

\begin{figure} [H]
\begin{center}
\includegraphics{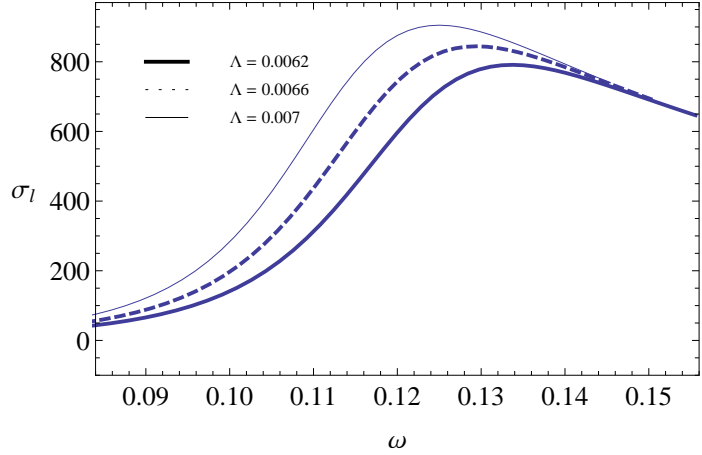}
\caption{The figure shows  $\sigma_l(\omega)$ vs $\omega$ for $g=1, M = 3$ and $l=2$. The value of $\Lambda$ takes the values $0.0062,0.0066, 0.007$.}
\label{crosslambda}
 \end{center}
 \end{figure}

When the cosmological constant is varied, the transmission coefficients (or greybody factors) become larger as given in Fig.$\refb{translambda}$. This quality is known to be the case for bulk and brane decay of scalar fields for (4 +n) dimensional Schwarzschild-de Sitter black holes \cite{kanti4}. Reflection coefficients correspondingly decreases with $\Lambda$ as given in Fig$\refb{translambda}$. The partial absorption cross section also increases as $\Lambda$ goes up as given in Fig$\refb{crosslambda}$.

\begin{figure} [H]
\begin{center}
\includegraphics{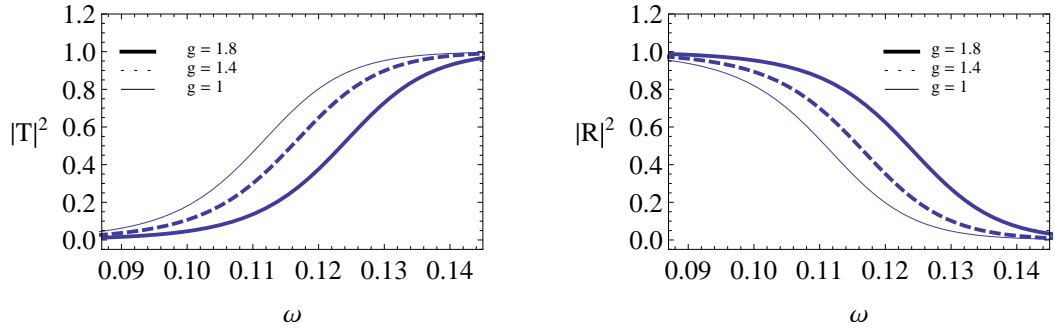}
\caption{The figure shows  $T(\omega)^2$ and $R(\omega)^2$ vs $\omega$ for $ \Lambda =0.007, M = 3$ and $l=2$. The value of $g$ takes the values $1, 1.4, 1.8$.}
\label{transq}
 \end{center}
 \end{figure}

\begin{figure} [H]
\begin{center}
\includegraphics{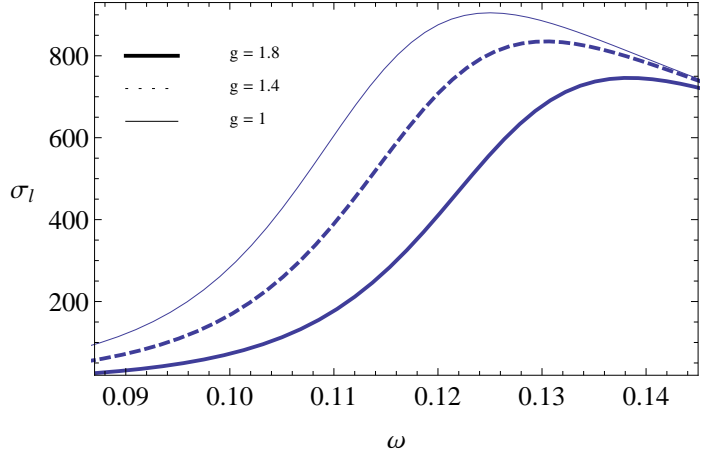}
\caption{The figure shows  $\sigma_l(\omega)^2$ vs $\omega$ for $ \Lambda =0.007, M = 3$ and $l=2$. The value of $g$ takes the values $1, 1.4, 1.8$.}
\label{crossq}
 \end{center}
 \end{figure}

When the magnetic charge $g$ is increased, the coefficient of transmission (greybody factors) decreases as in Fig$\refb{transq}$. Correspondingly, the reflection coefficient decreases with $g$. The partial absorption cross section is high  for low values of $g$ as given in Fig$\refb{crossq}$. Hence we notice that the absorption is less for high magnetic charge $g$ values. This is in fact is expected since the height of the effective potential in Fig$\refb{potg}$ increases when $g$ is increased. So there is less absorption for large $g$. Similar behavior was observed for the Bardeen black hole with $\Lambda=0$ in the paper by Macedo  and Crispino in \cite{macedo}.

\begin{figure} [H]
\begin{center}
\includegraphics{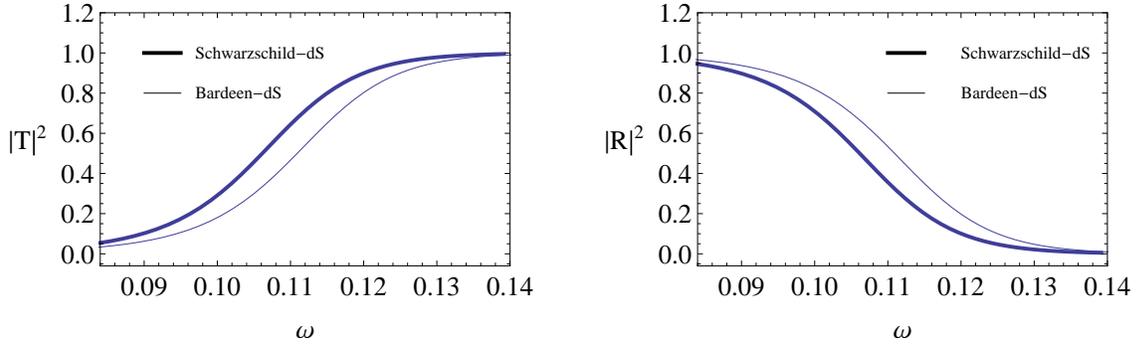}
\caption{The figure shows  $T(\omega)^2$  and $R(\omega)^2$ vs $\omega$ for the Schwarzschild-de Sitter and the Bardeen-de Sitter black holes. Here $ \Lambda =0.007, M = 3, l=2$ and $g =1$.}
\label{transboth}
 \end{center}
 \end{figure}

\begin{figure} [H]
\begin{center}
\includegraphics{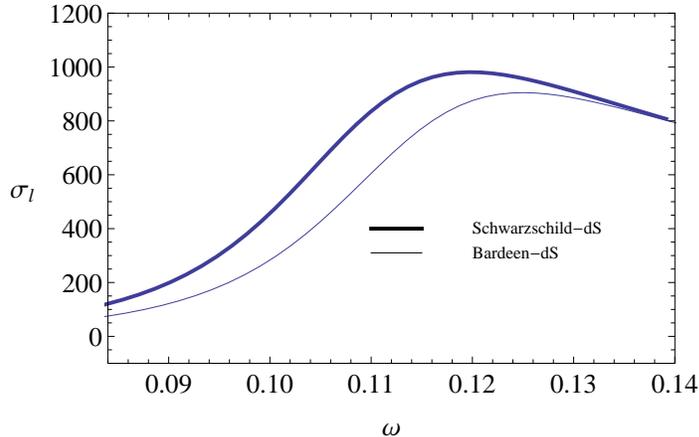}
\caption{The figure shows  $\sigma_l$ vs $\omega$ for the Schwarzschild-de Sitter and the Bardeen-de Sitter black holes. Here $ \Lambda =0.007, M = 3, l=2$ and $g =1$.}
\label{crossboth}
 \end{center}
 \end{figure}

We have also calculated $R$, $T$ and $\sigma_l$ for the Schwarzschild-de Sitter and Bardeen-de Sitter black hole for the same values of $\Lambda$, $M$. As shown in Fig$\refb{transboth}$,  Bardeen-de Sitter has a lower value  for $T$ and higher value for $R$ in comparison with the Schwarzschild-de Sitter black hole. The partial absorption cross section is lower for the Bardeen-de Sitter black hole as given in Fig$\refb{crossboth}$. This is in fact in agreement with the effective potentials given in Fig$\refb{potboth}$ where the height of the Bardeen-de Sitter potential is higher leading to lower absorption.

\section{Discussion}

In this paper we have presented a new class of black hole solutions which are regular; it is an extension of the well known Bardeen black hole with a positive cosmological constant. Bardeen dS black hole is regular everywhere as it's counterpart with a $\Lambda =0$. Such black holes could have three horizons depending on the parameters of the black holes such as $M$, $\Lambda$ and the magnetic monopole charge $g$. We have compared the horizons of the Bardeen dS black hole with the Schwarzschild dS black hole and observed that the event horizon is smaller for the Bardeen dS black hole.

We have discussed the various geometries for the degenerate black holes such as Nariai, cold and ultracold black holes.

The main goal of this paper is to present the scattering information of a massless scalar wave impinged on the Bardeen dS black hole. The effective potentials for the scalar wave is presented for various values of the parameters such as $l$, $g$ and $\Lambda$.  When $l$ and $g$ increases, the height of the potentials increase. We also observed that for $l=0$, the potential has a local minimum between the two horizons which is different compared to other potentials. When $\Lambda$ is increased, the height of the potential decreases. Schwarzschild dS black hole has a smaller potential than the Bardeen dS black hole.

To calculate the scattering information such as the transmission coefficient ($T$), reflection coefficient ($R$) and the partial cross section $\sigma_l$, we have employed the third order WKB approach. When the cosmological constant is increased, $T$ (or the grey body factor) and $\sigma_l$ increases. Hence the presence of  $\Lambda$ enhances the grey body factor and the absorption. When the magnetic charge $g$ is increased, $T$ and $\sigma_l$ decreases. Hence the presence of the magnetic charge decreases the absorption. Similar results have been reported for the the Bardeen black hole with $\Lambda =0$. We also compared the values for $\sigma_l$ and $T$ for the Schwarzschild dS and the Bardeen dS black hole. 

In continuing this work, it would be interesting to do a comparison of grey body factors of Reissner-Nordstrom-de Sitter black hole  and the Bardeen de Sitter black hole with the same charge. Also, computation of the quasi normal modes for this black hole for the spin 0, $\frac{1}{2}$ and $2$ would be  interesting. In this work, we have  not computed the $l=0$ case since the potential has a local minimum and the WKB approach is not effective. Another numerical method, such as the Runge-Kutta method would be appropriate to calculate cross sections for such a case.


\vspace{0.3 cm}

{\bf Acknowledgements:} SF like to thank  A.  Zhidenko   for providing the {\it Mathematica} file for the WKB approximation.


\end{document}